\def \SAIT #1 #2 {{\em Mem.\ Soc.\ Astron.\ It.\/} {\bf #1}, #2}
\def \MESS #1 #2 {{\em The Messenger\/} {\bf #1}, #2}
\def \ASTRNACH #1 #2 {{\em Astron. Nach.\/} {\bf #1}, #2}
\def \AAP #1 #2 {{\em Astron. Astrophys.\/} {\bf #1}, #2}
\def \AAL #1 #2 {{\em Astron. Astrophys. Lett.\/} {\bf #1}, L#2}
\def \AAR #1 #2 {{\em Astron. Astrophys. Rev.\/} {\bf #1}, #2}
\def \AAS #1 #2 {{\em Astron. Astrophys. Suppl. Ser.\/} {\bf #1}, #2}
\def \AJ #1 #2 {{\em Astron. J.\/} {\bf #1}, #2}
\def \ANNREV #1 #2 {{\em Ann. Rev. Astron. Astrophys.\/} {\bf #1}, #2}
\def \APJ #1 #2 {{\em Astrophys. J.\/} {\bf #1}, #2}
\def \APJL #1 #2 {{\em Astrophys. J. Lett.\/} {\bf #1}, L#2}
\def \APJS #1 #2 {{\em Astrophys. J. Suppl.\/} {\bf #1}, #2}
\def \APSS #1 #2 {{\em Astrophys. Space Sci.\/} {\bf #1}, #2}
\def \ASR #1 #2 {{\em Adv. Space Res.\/} {\bf #1}, #2}
\def \BAIC #1 #2 {{\em Bull. Astron. Inst. Czechosl.\/} {\bf #1}, #2}
\def \JSQRT #1 #2 {{\em J. Quant. Spectrosc. Radiat. Transfer\/} {\bf #1}, #2}
\def \MN #1 #2 {{\em Mon. Not. R. Astr. Soc.\/} {\bf #1}, #2}
\def \MEM #1 #2 {{\em Mem. R. Astr. Soc.\/} {\bf #1}, #2}
\def \PLR #1 #2 {{\em Phys. Lett. Rev.\/} {\bf #1}, #2}
\def \PASJ #1 #2 {{\em Publ. Astron. Soc. Japan\/} {\bf #1}, #2}
\def \PASP #1 #2 {{\em Publ. Astr. Soc. Pacific\/} {\bf #1}, #2}
\def \NAT #1 #2 {{\em Nature\/} {\bf #1}, #2}
\def\proptosima{$\; \buildrel \propto \over \sim \;$}
\def\proptosim{\lower.5ex\hbox{\proptosima}}            
\def\ltsima{$\; \buildrel < \over \sim \;$}
\def\simlt{\lower.5ex\hbox{\ltsima}}            
\def\gtsima{$\; \buildrel > \over \sim \;$}
\def\simgt{\lower.5ex\hbox{\gtsima}}            
\def\ea{et al.\ }

\documentstyle[twoside]{memsait}
\begin{opening}
\title{UNIFIED SCHEMES FOR RADIO-LOUD AGN:\\ 
RECENT RESULTS$^1$}
\author{PAOLO PADOVANI}
\institute{Dipartimento di Fisica, II Universit\`a di Roma ``Tor 
Vergata'', Roma, Italy}
\date{} 
\end{opening}

\begin{document}

\oddpagefooter{}{}{} 
\evenpagefooter{}{}{} 
\addtocounter{footnote}{1}
\footnotetext{Invited Review, to appear in {\it From the Micro- to the 
Mega-Parsec}, II Italian Conference on AGN} 

\medskip

\begin{abstract}
After briefly summarizing the main tenets of unified schemes of Active
Galactic Nuclei, I review some recent results in the field of unification of
radio-loud sources, both for the low-luminosity (BL Lacs and Fanaroff-Riley
type I radio galaxies) and high-luminosity (radio quasars and Fanaroff-Riley
type II radio galaxies) populations. 

\end{abstract}

\section{Introduction}

It now seems well established that the appearance of Active Galactic Nuclei
(AGN) depends strongly on orientation. Classes of apparently different AGN
might actually be intrinsically similar, only seen at different angles with
respect to the line of sight. The basic idea, based on a variety of
observations and summarized in Figure 1 of Urry \& Padovani (1995), is that
emission in the inner parts of AGN is highly anisotropic. The current paradigm
for AGN includes a central engine, surrounded by an accretion disk and by
fast-moving clouds, probably under the influence of the strong gravitational
field, emitting Doppler-broadened lines. More distant clouds emit narrower
lines. Absorbing material in some flattened configuration (usually idealized
as a torus) obscures the central parts, so that for transverse lines of sight
only the narrow-line emitting clouds are seen (narrow-lined or Type 2 AGN),
whereas the near-IR to soft-X-ray nuclear continuum and broad-lines are
visible only when viewed face-on (broad-lined or Type 1 AGN). In radio-loud
objects we have the additional presence of a relativistic jet, roughly
perpendicular to the disk, which produces strong anisotropy and amplification
of the continuum emission (``relativistic beaming''). In general, different
components are dominant at different wavelengths. Namely, the jet dominates at
radio and $\gamma$-ray frequencies (although it does contribute to the
emission in other bands as well), the accretion disk is thought to be a strong
optical/UV/soft X-ray emitter, while the absorbing material will emit
predominantly in the IR. It then follows that a proper understanding of AGN
will only come through multifrequency studies (e.g., Padovani 1997). 

This axisymmetric model of AGN implies widely different observational properties  
(and therefore classifications) at different aspect angles. Hence the need for
``Unified Schemes'' which look at intrinsic, isotropic properties, to unify
fundamentally identical (but apparently different) classes of AGN. Seyfert 2
galaxies have been ``unified'' with Seyfert 1 galaxies (see Antonucci 1993,
and references therein, and Granato, these Proceedings, for more recent
results). As regards the radio-loud population, Fanaroff-Riley type I (FR I:
i.e., low-power) radio galaxies have been unified with BL Lacertae objects, a
class of AGN characterized by very weak emission lines, while Fanaroff-Riley
type II (FR II: i.e., high-power) radio galaxies have been unified with radio
quasars. In the latter case, flat-spectrum radio quasars (FSRQ) are thought to
be oriented at relatively small angles w.r.t. to the line of sight ($\theta
\simlt 15^{\circ}$), while steep-spectrum radio quasars (SSRQ) should be at
angles intermediate between those of FSRQ and FR II radio galaxies. 

This paper is {\it not} meant to be a review of unified schemes: the
interested reader might consult, for example, the reviews by Antonucci (1993)
and Urry \& Padovani (1995). My aim is to discuss some of the very recent
results in the field of unified schemes for radio-loud AGN which have 
appeared in the literature in the last year or so (and therefore were not
included in Urry \& Padovani 1995). 

\section{BL Lac - FR I Unification}

\subsection{Broad lines in BL Lac (or ``When is BL Lac not a BL Lac?'')}

Vermeulen \ea (1995) have discovered a broad H$\alpha$ line in BL Lacertae,
the class prototype itself, comparable in luminosity and velocity width to 
that of the Seyfert 1 galaxy NGC 4151. Although weak, broad emission lines
have been previously detected in other BL Lacs (Stickel \ea 1993), the
presence of such a line in BL Lacertae itself has attracted the attention of
many BL Lac pundits. The main distinguishing feature of BL Lacs is in fact
their weak, or even absent, emission lines, generally characterized by
equivalent widths for the stronger lines $W_{\lambda} < 5$~\AA. The question
then is: Is BL Lac still a BL Lac? The reassuring answer is: Yes. The newly
detected broad line, in fact, has $W_{H\alpha} = 5.6\pm1.4$~\AA, consistent
with the 5 \AA~limit. 

This detection, however, has some quite interesting implications for our
understanding of BL Lacs (and of AGN in general: see Salvati, these
Proceedings). This relatively strong line should have been detected in
previous spectra, taken when the source was of roughly similar brightness.
Corbett \ea (1996) have estimated that the H$\alpha$ luminosity (and
equivalent width) must have increased by a factor of 4 from earlier
observations. If the source of ionization were related to the non-thermal
continuum, the equivalent width of all emission lines should be independent of
the ionizing power, as both line and continuum luminosity would be
proportional to it. Corbett \ea (1996) then suggest the presence of an
additional, variable continuum source: the observed $W_{\rm H\alpha}$ could in
fact be explained by an accretion disk whose flux is only 4\% of the
non-thermal, beamed continuum at the wavelength of H$\alpha$. Such a small
contribution to the total flux in the optical band affects only slightly the
shape of the optical continuum by flattening the slope in the 3,500 -- 7,000
\AA~range, but should produce a marked depolarization of the synchrotron
component (assuming thermal emission is unpolarized). This is a strong
prediction of the model although, as Corbett \ea point out, this dilution
might be difficult to detect due to the strong variability of the wavelength
dependence of the synchrotron polarization observed in BL Lacs. 

The possibility that BL Lacs possess an accretion disk, even though its
contribution to the total flux might be quite small, is suggestive.
Furthermore, the presence of broad lines in BL Lacs, and the lack of them (to
the best of my knowledge) in FR Is, might provide the only evidence, so far,
of the presence of an obscuring torus in low-luminosity radio sources.
Finally, the detection of an emission line at the same redshift as the host
galaxy absorption lines shows that at least BL Lac (and all the other objects
sharing this property: see e.g., Stickel \ea 1993) is not gravitationally
micro-lensed (Urry \& Padovani 1995). 

\subsection{BL Lacs (or lack thereof) in Abell Clusters}

Owen, Ledlow \& Keel (1996) have looked for BL Lacs in 183 low-luminosity radio
galaxies in relatively rich ($\sim 60\%$ of richness class 1 and 2) clusters at
$z < 0.09$. They found no object which met the ``standard definition'' of BL
Lacs, based on a measure of a non-thermal component (4000 \AA~break $\simlt
25\%$) and weakness of the lines ($W_{\lambda} < 5$~\AA), although they found
4 objects with weak evidence for non-thermal activity (4000 \AA~break $\sim 33
- 43\%$: typical ellipticals have break values $\sim 50\%$), two of them with
relatively strong lines (unlike BL Lacs). Owen \ea expected 8 BL Lacs on the
basis of a relativistic beaming model (Urry, Padovani \& Stickel 1991) or 3/18
at $P > 10^{32}$ erg s$^{-1}$ Hz$^{-1}$ simply on the basis of the observed
luminosity functions of FR Is and BL Lacs. Their null results are inconsistent
with these expectations at the $> 99\%$ and $95\%$ level respectively,
apparently putting in serious trouble the BL Lac -- FR I unification. This
inconsistency might be at least partly related to the definition of a BL Lac,
which is likely to be not that clear-cut and should perhaps be slightly
relaxed, for example by increasing the value of the 4000 \AA~break (as pointed
out by March\~a \ea 1996). More importantly, it seems now quite likely that
BL Lacs avoid the richest clusters (at least at low redshifts: e.g., Pesce,
Falomo \& Treves 1995; Smith, O'Dea \& Baum 1995; Wurtz \ea 1996b) and reside
mostly in relatively poor (richness class $\sim 0$) clusters, which however
seem to include a large fraction of FR I galaxies (Hill \& Lilly 1991). This 
behaviour, although still unexplained, is nevertheless consistent with the
findings of Wurtz, Stocke \& Yee (1996a) that most BL Lacs have host galaxies
too faint to be bright cluster galaxies (BCGs) in rich clusters: their host
galaxy magnitudes are more typical of BCGs in poorer clusters, which are less
luminous. 

In summary, the sample of Owen \ea (1996) probably included a large number of
unlikely BL Lac hosts: the significance of their null results should then be
re-evaluated. 

\subsection{X-ray selected vs. Radio-selected BL Lacs (or HBL vs.
LBL)$^2$}

\addtocounter{footnote}{1}
\footnotetext{This is a big topic, currently hotly debated in the literature,
which is impossible to address in detail in such a short contribution.
Therefore, after introducing the problem, I will simply refer to some recent
papers on the subject.} 

It has long been recognized that BL Lacs appear to come in two versions, one
being more extreme, having larger optical polarization, core-dominance, and
variability, and usually associated with objects selected in radio surveys, the
so-called radio-selected BL Lacs (RBL). The less extreme version, mostly
selected in X-ray surveys, was then obviously named accordingly
(X-ray-selected BL Lacs: XBL). This division was clearly not satisfactory, as
it was not based on intrinsic physical properties but solely on the selection
band, and did not uniquely characterize a source. Recent all-sky surveys, in
fact, include many BL Lacs previously selected at radio frequencies and which
could then now be classified both as RBL {\it and} XBL. Padovani \& Giommi
(1995; see also Ledden \& O'Dell 1985 for the first subdivision of BL Lac 
classes based on $L_{\rm x}/L_{\rm r}$) have then proposed to classify BL Lacs
in HBL (or high-energy peak BL Lacs) and LBL (or low-energy peak BL Lacs),
based on a simple division in terms of their X-ray-to-radio flux ratio. This
reflects their different multifrequency spectra, with LBL (most RBL) having a
peak (i.e., emitting the bulk of their energy) at infrared-optical
frequencies, and HBL (most XBL) having a peak at ultraviolet-X-ray
frequencies. Note that such a distinction does not mean that there are two
separate BL Lac classes: it is more likely that there is a continuous
distribution of, for example, peak frequency, which turns out to be bimodal
because of selection effects. 

What is the relation between the two classes? This is being (and has been)
discussed by various authors (see Padovani \& Giommi 1995 and Urry \& Padovani
1995, and references therein). Briefly, the two opposite views are that: 1.
LBL are {\it intrinsically} more numerous (``different energy distribution''
scenario); 2. HBL are {\it intrinsically} more numerous (``different angle''
scenario). In the former view, radio selection gives an ``unbiased'' view of
the BL Lac population (in terms of population ratios) and HBL and LBL are
oriented w.r.t. the line of sight within approximately the same angles. In the
latter, X-ray selection is unbiased, and HBL are seen at larger angles.
Currently available samples are not deep enough to overthrow the balance in
favour of one or the other of the two explanations, which diverge in their
predictions only at relatively faint radio and X-ray fluxes. 

Some of the recent papers on the subject include: Padovani \& Giommi (1996) and
Lamer, Brunner \& Staubert (1996), on the different {\it ROSAT} spectra of HBL
and LBL, which are consistent with their different multifrequency spectra;
Sambruna, Maraschi \& Urry (1996) on, amongst other things, the difficulty in
explaining the observed differences in the multifrequency spectra of HBL and 
LBL only by changing viewing angle; Kollgaard \ea (1996) on radio constraints
to the two pictures, which would seem to favour the ``different angle'' 
hypothesis. Finally, Wurtz \ea (1996a) find no differences in the host galaxy 
properties of HBL and LBL, while Fossati et al. (these Proceedings), present 
yet a new explanation for the relation between HBL and LBL. 

\section{Radio Quasar - FR II Unification} 

\subsection{Linear sizes of radio galaxies and quasars}

The question of the comparison between linear sizes of radio galaxies and
quasars lies at the core of unified schemes: if radio quasars are oriented at
small angles to the line of sight, then they should have systematically
smaller large-scale radio structures than radio galaxies, thought to be in the
plane of the sky. A simple relation exists between the ratio of median sizes
of radio galaxies and quasars, $r_{\theta}$, and the half-opening angle of the
obscuring torus (or equivalently the critical angle w.r.t. the line of sight
separating the two classes), $\theta_{\rm c}$. (In fact, the smaller
$\theta_{\rm c}$, the higher will be $r_{\theta}$.) The critical angle is also
obviously related to the quasar fraction $f_{\rm q}$ ($f_{\rm q} = 1 - \cos
\theta_{\rm c}$), from which it follows that $r_{\theta}$ and $f_{\rm q}$ have
to be inversely related (that is the smaller $f_{\rm q}$, the smaller
$\theta_{\rm c}$, the higher $r_{\theta}$). The important point of Barthel's
(1989) paper was to show that both $r_{\theta}$ and $f_{\rm q}$ were
consistent with a critical angle $\simeq 45^{\circ}$ for the 3C sample at $0.5 <
z < 1.0$. Barthel's findings were criticized by some authors, particularly by
Singal (1993), who argued that not only the quasar fraction seemed to be
redshift dependent but the apparent agreement between the angles inferred from
the number ratios and those derived from the linear sizes disappeared
extending the comparison to other redshift ranges. 

It must be stressed that the whole subject of linear sizes of radio sources
is quite complex, having been addressed by many authors quite often with
contrasting results (see Neeser et al. 1995 for a very good description of
some of the selection effects which can affect this sort of analysis). As
summarized in Urry \& Padovani (1995) and discussed in Singal (1996), various
solutions to the problems raised by Singal (1993) have been proposed. Singal
(1996), however, examining the quasar fractions observed in low-frequency
radio samples, noted that not only the quasar fraction decreases at lower
fluxes (which would show that a single $\theta_{\rm c}$ is untenable) but the
ratio of median sizes of radio galaxies and quasars actually seems to {\it
decrease} as well, contrary to the expectations of unified schemes. In other
words, while $r_{\theta}$ should be inversely dependent on $f_{\rm q}$,
$r_{\theta}$ actually {\it increases} with $f_{\rm q}$ (see Figure 2 of Singal
1996). Note that no change of $\theta_{\rm c}$ with redshift or power will
bring the data in agreement with the predictions of unified schemes, as the
observed dependence is opposite to the predicted one. 

Does this mean that unified schemes are totally wrong? Apparently not, if one
takes into account the evolution of the single radio sources. Recent results
(e.g., Fanti \ea 1995; Readhead \ea 1996) suggest that radio sources grow in
size during their active phase, with an accompanying decrease in power due to 
expansion losses. Gopal-Krishna, Kulkarni \& Wiita (1996) have parameterized 
this behaviour, with the further assumption that the half-opening angle of the 
torus depends on luminosity (which could be explained by the fact that more
powerful engines ``erode'' away the inner parts of the torus: see Urry \&
Padovani 1995 and references therein). The result is that Gopal-Krishna \ea
can explain the apparently anomalous relation between $r_{\theta}$ and $f_{\rm
q}$ pointed out by Singal (1996) with reasonable input parameters. Within this
model, the median radio sizes of quasars can actually approach, or even exceed,
that of radio galaxies, as reported by Singal (1996) for some samples. 
 
Note that the inclusion of source evolution in unified schemes, which seems
unavoidable, might complicate the use of the core-to-extended flux ratio, $R$,
as an orientation indicator, as in this view the extended flux would also be
time-dependent, that is $R$ could change with time for the same angle to the
line of sight. 

\subsection{Infrared spectroscopy of radio galaxies}

There are some ways to overcome the presence of obscuring material and be able
to see the central engine even in Type 2 sources. One of these is
spectropolarimetry, where one looks for broad lines in that small fraction of
the nuclear flux which is in some cases ``reflected'' towards our line of
sight by dust and/or electrons. Recent results in this field are discussed by
Cimatti, these Proceedings. I concentrate here on another way to see through
the dust by moving to longer wavelengths, where the torus becomes more
transparent: infrared spectroscopy. Infrared broad lines have been detected
before in Type 2 radio sources (see Urry \& Padovani 1995 for references) but
it is clear that one would like to be able to answer such questions as: How
many Type 2 sources actually show broad lines when observed in the infrared?
What is the distribution of BLR extinction in these sources? The best way to
address these points is through the use of an unbiased, flux-limited sample,
such as that studied by Hill, Goodrich \& DePoy (1996). These authors present
infrared spectrophotometry of Pa$\alpha$, plus optical spectroscopy of
H$\alpha$ and H$\beta$, for a flux-limited sample of 11 3CR FR II radio
sources with $0.1 \le z < 0.2$, which includes 8 narrow-line radio galaxies, 2
broad-line radio galaxies, and one quasar. Infrared broad lines are detected
in 3 narrow-line radio galaxies (or $\sim 40\%$ of the Type 2 sources), and
extinction values are derived for narrow and broad lines. The distribution of
broad-line extinction, $A_{\rm V}$, is in reasonable agreement with a simple
unification model. Extension of this work to higher redshifts (and therefore
luminosities) should permit a direct test of the idea that the half-opening
angle of the torus increases with power (see Sect. 3.1). If that is the case,
at higher powers there should be fewer sources with relatively large
extinctions ($A_{\rm V} \simgt 5$), as compared to their lower power
counterparts. 

\section{Across the two Schemes}

\subsection{Relation between BL Lacs and Flat-spectrum radio quasars}

The relation between BL Lacs and FSRQ is not clear. The two classes share
similar properties (apart from the presence of strong emission lines in the
latter class). Sambruna \ea (1996) have suggested a continuity in the
multifrequency spectra of HBL, LBL, and FSRQ, with the frequency of the peak
of the emission shifting to higher values for decreasing luminosities. This
continuity seems to have been severely weakened by the discovery of
``HBL-like'' quasars (Padovani, Giommi \& Fiore, these Proceedings), that is
of radio quasars (including some FSRQ) with multifrequency spectra similar to
those of HBL. Padovani, Giommi \& Fiore (1997) have also found that, contrary
to previous results, even the soft X-ray spectra of FSRQ and BL Lacs of the
LBL type are similar. Finally, Vagnetti (these Proceedings), has presented his
latest results on the possible evolutionary relation between the two classes. 

\section{Open Questions - Hot Topics and Conclusions}

I have listed here some of the open questions/hot topics on which work is in 
progress. These will be hopefully discussed at the next Italian meeting on 
AGN, to be held in 1998!  

\begin{itemize}

\item Obscuration in FR Is. The presence of broad lines in at least some BL 
Lacs (Sect. 2.1) indicates that some obscuring material must be present in
FR Is (as well as in FR IIs). A SAX core program will address this issue.
Also, ISO observations of FR Is should provide further constraints. 

\item BL Lac (and FR I) environment. As stressed in Section 2.2, it might well 
be that BL Lacs avoid the richest clusters. More work in this field is needed,
as the samples used to study the environment of BL Lacs are still relatively 
small. Various ongoing projects will address this point in more detail.
Comparable studies on the environment of FR I radio galaxies using sizeable
and well defined samples should also be performed. 

\item HBL/LBL dichotomy. As mentioned in Section 2.3, we need deeper and
larger BL Lac samples to address this point. Amongst the various groups
working on this, both Perlman, Padovani, Giommi, \ea (1997) and Caccianiga,
Maccacaro, Wolter, \ea (these Proceedings) are cross-correlating X-ray and
radio catalogues to select new BL Lac (and FSRQ) samples. 

\item (Spectro)Polarimetry of complete samples of radio-galaxies. As mentioned
in Section 3.2, spectropolarimetry allows one to see emission from the nuclear
region, particularly broad lines, even in Type 2 sources. So far, however,
spectropolarimetric studies have concentrated on individual, sometimes
peculiar, sources. We would need a spectropolarimetric study of a complete,
well-defined sample, to quantify how common reflected broad lines are, for
example. As far as I know, nobody is doing this, but I hope that I am 
misinformed on this! Tadhunter \ea and di Serego Alighieri et al., however,
are carrying out a polarimetric study of a complete sample of radio sources,
which should still give interesting, quantitative information. 

\item VLBI studies of complete sample of radio sources. These are important to
study, among other properties, the distribution of superluminal speeds to
estimate beaming parameters and constrain unification. Various surveys are in
progress, for example the Caltech-Jodrell Bank survey of strong ($0.7 < f_{\rm
6cm} < 1.3$ Jy) radio sources (e.g., Polatidis \ea 1995) or the survey of
low-power radio galaxies by the Bologna group (e.g., Venturi \ea 1995). 

\end{itemize}

In conclusion, the unification of radio-loud sources is a very active field,
with interesting results appearing even on a relatively short time scale. In
1995 alone, about 65 papers on beaming and unification of radio sources have
been published (51 in refereed journals), an all-time record judging from the
data relative to the last ten years (source: Astrophysics Data System).
Various modifications to our relatively simple and idealized unification
picture seem to be in order: not all FR I radio galaxies might host BL Lacs,
after all, but probably only those in relatively poor clusters; obscuring
material might be present in FR I radio galaxies as well, and not only in
their higher-luminosity counterparts; not unexpectedly, radio sources do
evolve in time, growing in size and dimming in power, which might help solving
some inconsistencies in their linear sizes relative to those of radio
galaxies; infrared broad lines are indeed detected in about half of the
low-redshift narrow line radio galaxies; but we still do not know if BL Lacs
of the LBL type are intrinsically more numerous than BL Lacs of the HBL type! 
 
Finally, although so far unified schemes for radio-loud and radio-quiet AGN
have been considered separately, there might be some connection between the
two. Falcke, Sherwood \& Patnaik (1996) have recently suggested that
relativistic boosting, normally associated with radio-loud AGN, might be
present in the radio cores of some radio-quiet sources as well. This idea,
with its far-reaching implications, might open up new research paths which
could help us to solve the long-standing problem of the radio-loud/radio-quiet 
dichotomy of the AGN population. 


\end{document}